\documentclass[conference]{IEEEtran}
\IEEEoverridecommandlockouts

\usepackage{cite}
\usepackage{amsmath,amssymb,amsfonts}
\usepackage{graphicx}
\usepackage{textcomp}
\usepackage{xcolor}
\usepackage{algorithm}
\usepackage{algpseudocode}
\usepackage{subfig}
\def\BibTeX{{\rm B\kern-.05em{\sc i\kern-.025em b}\kern-.08em
    T\kern-.1667em\lower.7ex\hbox{E}\kern-.125emX}}
\makeatletter
\newcommand{\linebreakand}{%
      \end{@IEEEauthorhalign}
      \hfill\mbox{}\par
      \mbox{}\hfill\begin{@IEEEauthorhalign}
}    
\makeatother
\begin{document}

\title{On deploying the Artificial Sport Trainer into practice}

\author{\IEEEauthorblockN{1\textsuperscript{st} Iztok Fister Jr.}
\IEEEauthorblockA{\textit{Faculty of Electrical Engineering and Computer Science} \\
\textit{University of Maribor}\\
SI-2000 Maribor, Slovenia}
\and
\IEEEauthorblockN{2\textsuperscript{nd} Iztok Fister}
\IEEEauthorblockA{\textit{Faculty of Electrical Engineering and Computer Science} \\
\textit{University of Maribor}\\
SI-2000 Maribor, Slovenia}
\linebreakand
\IEEEauthorblockN{3\textsuperscript{rd} Andres Iglesias}
\IEEEauthorblockA{\textit{Dpt. of Applied Mathematics and Computational Sciences} \\
\textit{University of Cantabria}\\
39005 Santander, Spain}
\and
\IEEEauthorblockN{4\textsuperscript{th} Akemi Galvez}
\IEEEauthorblockA{\textit{Dpt. of Applied Mathematics and Computational Sciences} \\
\textit{University of Cantabria}\\
39005 Santander, Spain}
\linebreakand
\IEEEauthorblockN{5\textsuperscript{th} Suash Deb}
\IEEEauthorblockA{\textit{IT \& educational Consultant} \\
\textit{Ranchi}\\
Jharkhand, India}
\and
\IEEEauthorblockN{6\textsuperscript{th} Du\v{s}an Fister}
\IEEEauthorblockA{\textit{Faculty of Economics and Business} \\
\textit{University of Maribor}\\
SI-2000 Maribor, Slovenia}
}

\maketitle

\begin{abstract}
Computational Intelligence methods for automatic generation of sport training plans in individual sport disciplines have achieved a mature phase. In order to confirm their added value, they have been deployed into practice. As a result, several methods have been developed for generating well formulated training plans on computers automatically that, typically, depend on the collection of past sport activities. However, monitoring the realization of the performed training sessions still represents a bottleneck in automating the process of sport training as a whole. The objective of this paper is to present a new low-cost and efficient embedded device for monitoring the realization of sport training sessions that is dedicated to monitor cycling training sessions. We designed and developed a new bike computer, i.e. the AST-Monitor, that can be mounted easily on almost every bicycle. The aforementioned bike computer is based on the Raspberry Pi device that supports different external sensors for capturing the data during the realization of sport training sessions. An adjusted GUI tailored to the needs of athletes is developed, along with the hardware. The proof of concept study, using the AST-Monitor in practice, revealed the potential of the proposed solution for monitoring of realized sport training sessions automatically. The new device also opens the door for the future utilization of Artificial Intelligence in a wide variety of sports. 
\end{abstract}


\section{Introduction}
Sport training is a complex task consisting of four phases~\cite{fister2019computational}: (1) planning, (2) realization, (3) control, and (4) evaluation. Many decades ago, sport trainers were left to the pencil and notebook when they were planning sport training sessions, as well as controlling the performance achieved during the realization by their athletes. All the training plans were written down carefully, aligned with the performances of different athletes who were in the domain of the sport trainer. Sport trainers in cycling, for instance, then compared the results on one particular track with the past results on the same track, and drew graphs of performances in their notebooks. The performance analysis which appeared in notebooks was one of the first data analyses in the realm of sport training. These analyses then offered a very effective way for adapting the sport training plans in the future.

Obviously, tracking and monitoring data during the realization of sport training sessions are just one side of the coin. Preprocessing the achieved data and using these data for intelligent performance analysis is the other side. With the huge numbers of different data science methods, we are able to build practically the whole models of sport training performances, together with future predictions, in order to enhance the performances of different athletes.

Use of smart wearable devices is now becoming inevitable during the process of sport training. Different smart gadgets, such as, for example, smart watches or smartphones, can be found on almost every runner running in the nearest park, or on every cyclist on the local roads. However, most of these devices are still too general, especially, in the case of personalized sport training planning. In the past, a lot of custom devices were built in the area of sport training. For example, Baca and Kornfeind~\cite{baca2006rapid} proposed feedback systems that were built of embedded sensors and integrated in sport equipment. Research suggested that these solutions should
help athletes in more efficient comparison of their performances. On the other hand, Novatchkov and Baca~\cite{novatchkov2013artificial} presented the potential of Artificial Intelligence (AI) methods in weight training. 
A MOPET~\cite{buttussi2008mopet} is an example of a wearable system aimed at supervising a physical fitness activity. Activity recognition is also one important domain where custom wearable devices are developed~\cite{bulling2014tutorial,velloso2013qualitative,zhou2016never,nguyen2015basketball}. 
 
Artificial Sport Trainer (AST), proposed by Fister et al.~\cite{fister2015computational}, is a relatively young intelligent system in sport that incorporates several data mining algorithms for searching the characteristics of athletes in training, as well as generating sport training sessions for them. The concept of the AST is utilized mostly on the theoretical level and in simulations. 

In this paper, we pay special attention to the realization phase of sport training sessions for cycling. We developed a custom embedded device that actually represents an interface between an athlete and the algorithm/method. The proposed device is called the AST-Monitor, which is basically a Raspberry Pi device equipped with different external sensors for monitoring real-time data. The special Graphical User Interface (GUI) enables the cyclist to interact with the device, and provides a feedback from the bicycle.

Altogether, the main contributions of this paper are summarized in the following points:
\begin{itemize}
    \item a new embedded device is built,
    \item a new GUI is developed for interaction with the cyclist during the training session,
    \item a practical utilization of the proposed system  is tested in the real-world,
    \item many goals and challenges are identified for future works.
\end{itemize}

The structure of the remainder of the paper is as follows: Section~\ref{sec:2} outlines the design of the proposed AST-Monitor. The implementation of the AST-monitor is the subject of Section~\ref{sec:3}. In Section~\ref{sec:4}, the proof of the proposed concept is tested, while the paper concludes with Section~\ref{sec:5}, where the performed work is summarized and the directions are outlined for the future work.

\section{Design of the AST-Monitor}\label{sec:2}
The motivation behind the development of the AST-Monitor was to build a mobile device that would be capable of supplementing the AST by supporting the realization phase of the sport training sessions. Although there are numerous smartphone applications and smartwatches able to track sport training sessions that enable post visualization of performed activities, and even go further with the predictions of training plan generation, no solution exists which could be integrated in all four phases of the sport training entirely on the personal level. 

\begin{figure*}[htb]
  \centering
    \includegraphics[width=0.9\textwidth]{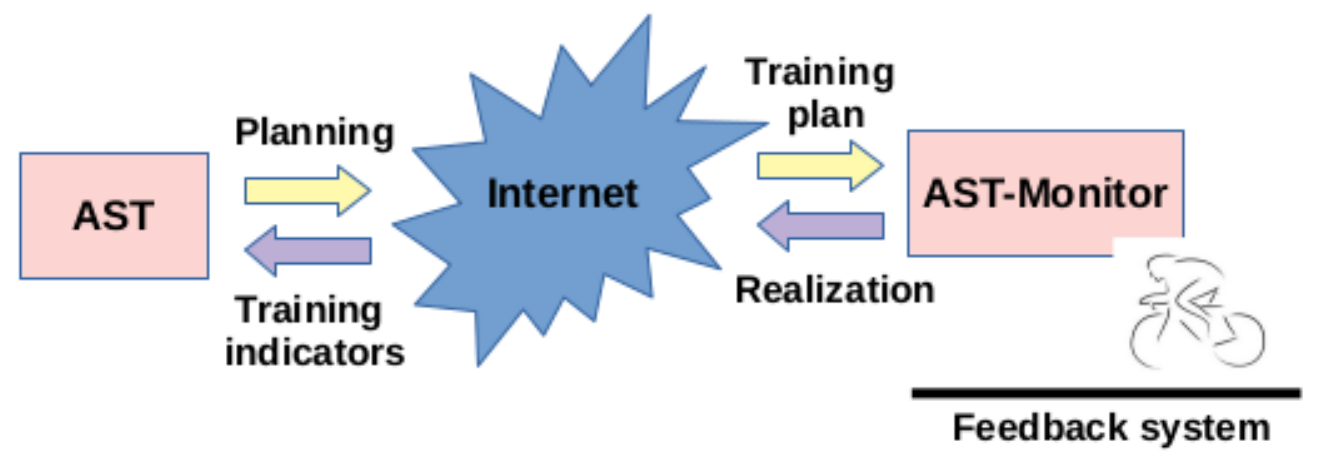}
    \caption{Design of the AST-Monitor.}
    \label{fig:design}
\end{figure*}

In line with this, the AST-Monitor can read training plans created by the AST, monitor their realization and control the performances of the realized training sessions by various training load indicators automatically. These indicators are forwarded to the AST for the detailed performance analysis after the athlete finishes the training session (Fig.~\ref{fig:design}).
Let us mention that the AST-Monitor is designed as a feedback system for cyclists, with the purpose of minimizing the difference between (1) prescribed and (2) measured Heart Rates (HRs) training load indicators as far as possible.

The designed AST-Monitor needs to accomplish the following goals:
\begin{itemize}
    \item the device must be small enough to be embedded on the bike, 
    \item the device must be flexible for adding different sensors measuring the additional training load indicators,
    \item the device must allow us to build a personalized GUI,
    \item the device must be able to run at least some computational intelligence 
    algorithms as demanded by the AST,
    \item the device must be able to connect to the Internet,
    \item the device must allow the presentation of the load indicators in graphical form, which is the easiest way to forward information to the cyclist,
    \item the device must give us complete access to the filesystem.
\end{itemize}
In our study, Raspberry Pi was found to be the best price/performance device for fulfilling all these goals.

\section{Implementation of the AST-Monitor}\label{sec:3}
The implementation of the AST-Monitor was indeed very complex, and was divided into three different parts:
\begin{itemize}
    \item hardware: connecting the hardware components together,
    \item software: design and implementation of the algorithms for supporting the feedback system,
    \item deployment: installing the hardware device on a bicycle.
\end{itemize}
In the remainder of the section, the mentioned parts are discussed in detail.

\subsection{AST-Monitor Hardware}
\begin{figure}
  \centering
    \includegraphics[width=0.5\textwidth]{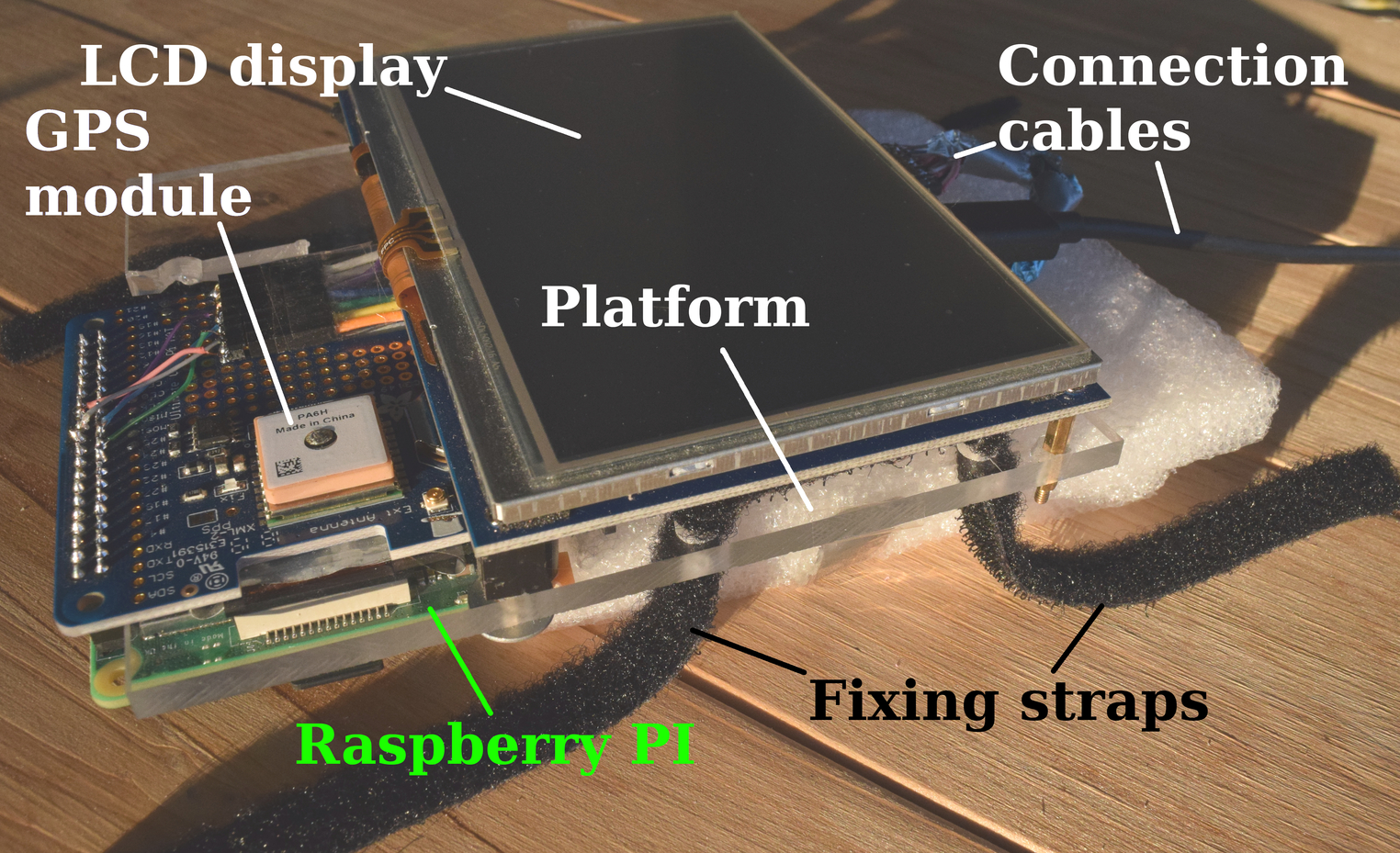}
    \caption{Hardware architecture of the AST-Monitor.}
    \label{fig:HW}
\end{figure}

The complete hardware part is shown in Fig.~\ref{fig:HW}, 
from which it can be seen that the AST-computer is split into the following pieces: 
\begin{itemize}
    \item a platform with fixing straps that attach to a bicycle,
    \item the Raspberry Pi 4 Model B micro-controller with Raspbian OS installed,
    \item a five-inch LCD touch screen display,
    \item a USB ANT stick,
    \item Adafruit's Ultimate GPS HAT module.
\end{itemize}
A Serial Peripheral Interface (SPI) protocol was dedicated for communication between the Raspberry Pi and the GPS peripheral. A specialized USB ANT stick was used to capture the HR signal. The screen display was connected using a modified (physically shortened) HDMI cable, while the touch feedback was implemented using physical wires. The computer was powered during the testing phase using the Trust's (5 VDC) power-bank. The AST-Monitor prototype is still a little bulky, but a more discrete solution is being searched for, including the sweat drainer of the AST.

\subsection{AST-Monitor Software}
The AST-Monitor software supports the so-called interval training sessions, where each exercise is repeated more times. Such training is characterized with combinations of highly intense periods (of typically shorter duration), after which rest phases follow. The software part consists of two algorithms:
\begin{itemize}
    \item an algorithm for realization of the training plan (Algorithm~\ref{alg:1}),
    \item an algorithm for feedback control (Algorithm~\ref{alg:2}).
\end{itemize}
The former is devoted for proper realization of the interval training exercises, while the latter for giving a feedback. Let us mention that the intensity phase of the interval training exercise, and, consequently, the algorithm for feedback control, is started by pressing the appropriate button on the AST-Monitor.
\begin{algorithm}
\small
\caption{Algorithm for training plan realization.}
\label{alg:1}
\begin{algorithmic}
\State $ID\leftarrow 0$;
\While {\Call{Training\_not\_finished}{}} \Comment{When $ID=\max ID$}
\State $\langle \overline{HR}_{ID},t_{ID}\rangle\leftarrow$\Call{Training\_plan}{$ID$}; \Comment{Search for exercise}
\State \Call{Wait\_for\_start\_button}{}; \Comment{Start by pressing button}
\State \Call{Feedback\_control}{$\langle \overline{HR}_{ID},t_{ID}\rangle$}; \Comment{Manipulating HW}
\State $ID\leftarrow ID+1$;
\EndWhile
\end{algorithmic}
\end{algorithm}

Algorithm~\ref{alg:2} continues for the duration of the training session, with each second (1) receiving the value of the HR training load indicator, (2) calculating the new average value $\overline{\mathit{HR}}$, and \begin{algorithm}
\small
\caption{Algorithm for feedback control.}
\label{alg:2}
\begin{algorithmic}
\Require $\overline{HR}_{ID},~t_{ID}$

\State $t\leftarrow 0$; \Comment{Elapsed time of cycling}
\State $n\leftarrow 1$; \Comment{Times of sampling}
\State $\overline{HR}\leftarrow 0$; \Comment{Average $HR$ calculation}
\While{$t\leq t_{ID}$}
\State $\langle HR,t \rangle\leftarrow$\Call{Read\_control\_data}{}; \Comment{Obtain sample data}
\State $\overline{HR}\leftarrow \overline{HR}+\frac{1}{n}\left(HR-\overline{HR}\right)$; \Comment{Calculate average $HR$}
\State \Call{Display\_control\_data}{$n,\overline{HR},t,\overline{HR}_{ID},t_{ID}$}; \Comment{Update}
\EndWhile
\end{algorithmic}
\end{algorithm}
(3) refreshing the feedback display data for the cyclist, who can on the basis of this feedback adapt the intensity of the ride.

\subsection{Deploying the AST-Monitor}
\begin{figure*}[htb]
\begin{minipage}{.496\linewidth}
  \includegraphics[width=\linewidth]{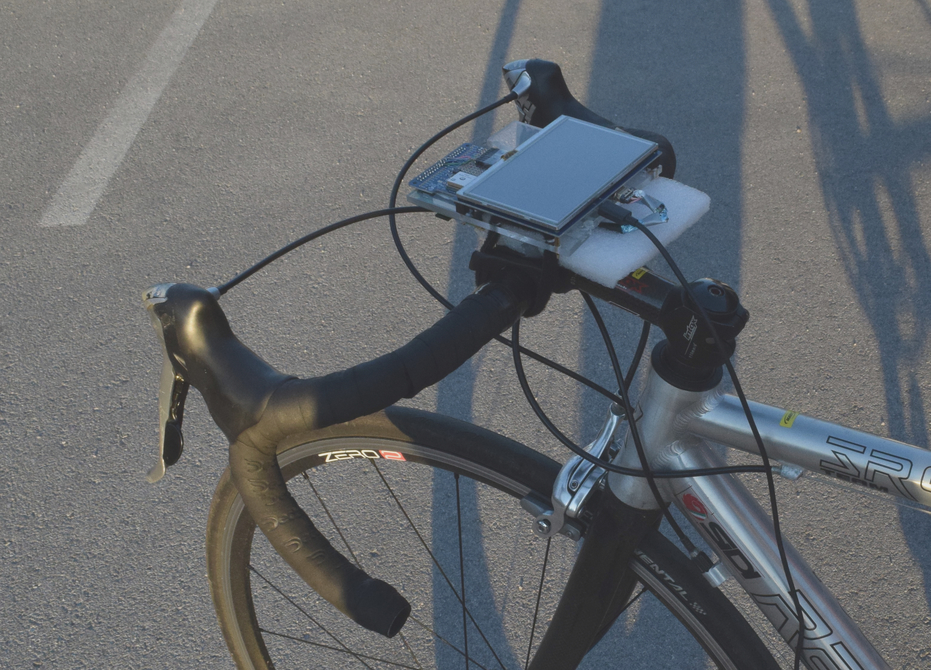}
  \label{fig:img2}
\end{minipage}
\begin{minipage}{.504\linewidth}
  \includegraphics[width=\linewidth]{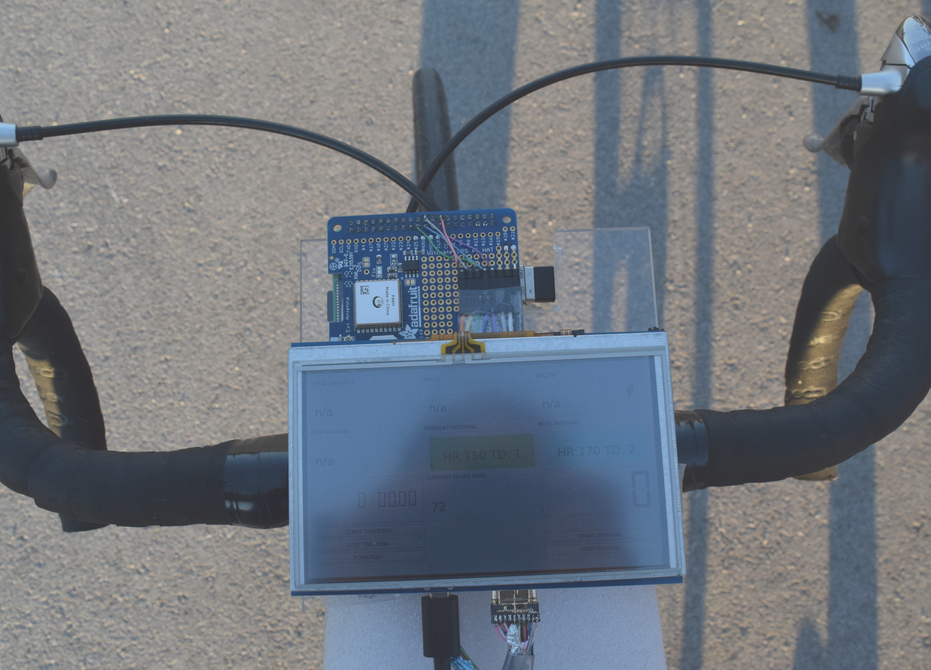}
  \label{fig:img2}
\end{minipage}
\caption{Deployment of the AST-Monitor on a real cycle.}
\label{fig:deploy}
\end{figure*}

In order to use the AST-Monitor as a wearable device, its hardware part should be mounted on a bicycle to follow the cyclist at all times. In case that cyclist becomes distant from the AST-Monitor, the wireless signal could be lost and connection needs to be re-established afterwards.  
Fig.~\ref{fig:deploy} illustrates the position of the AST-Monitor on a bicycle. The figure is divided into two subfigures, where Fig.~\ref{fig:deploy}a shows the proposed wearable device from an engineering point of view, while Fig.~\ref{fig:deploy}b from the user's (i.e., the athlete's) one.

From an engineering point of view, it can be seen that the AST-computer is an integral, unified and freely customizable bike computer. As it is supposed to monitor and give critical information about the cyclist's performance in real-time, it was designed to attach easily on the top of the bicycle's handlebars. A specialized plexiglas platform with a polystyrene cushion were devoted for stabilization of the AST-computer, while two fixing straps were used for attachment to the handlebars. Prior to attachment, the cycle handlebars were equipped with two separate carriers.

\section{Illustrative example}\label{sec:4}
The purpose of our experimental work was to show that the concept of the AST-Monitor works, and that it is appropriate for use in practice. In line with this, the proposed wearable device was tested in a simulated environment, where we focused on particular functionalities of the solution. A training plan for cycling consisting of interval training sessions was selected in our study, due to the short duration and repeating of the same exercises (Table~\ref{tab:1}). Thus, each exercise consisted of an intensity phase followed by a rest phase. All data in the table refer to the intensity phase.
\begin{table}[htb]
\caption{Definition of the plan for interval training.}
\label{tab:1}
\centering
\begin{tabular}{ c c c }
ID & $\overline{HR}_{ID}$ & $t_{ID}\text{[min]}$  \\
\hline
1 & 150 & 1 \\ 
2 & 170 & 2 \\  
3 & 145 & 2 \\    
4 & 180 & 1 \\    
5 & 182 & 1   
\end{tabular}
\end{table}

\begin{figure*}[htb]
  \centering
    \includegraphics[width=0.8\textwidth]{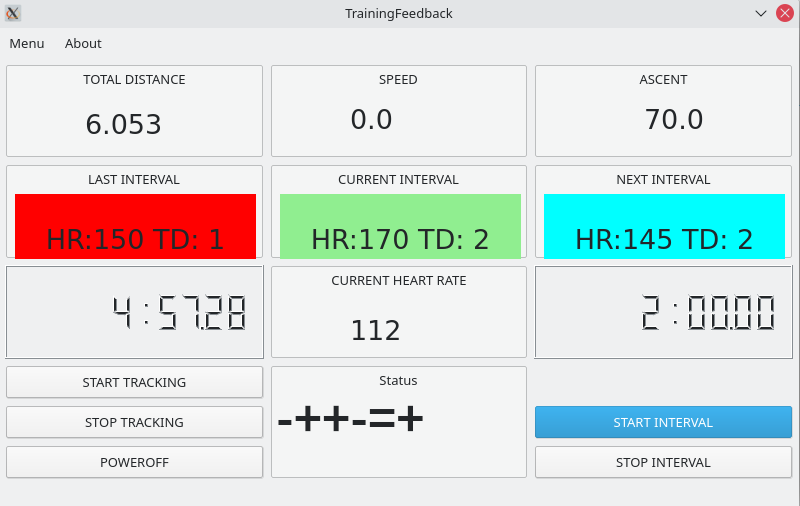}
    \caption{GUI application (used in the simulation environment.)}
    \label{gui-app}
\end{figure*}

The GUI enabled users to communicate with the proposed device. This was developed fully in the Python programming language using the PyQT library. The GUI was adapted to work consistently with a 5 inch display. Fig.~\ref{gui-app} presents the outlook of the developed application. 
Basically, the GUI consists of four segments. The first segment presents the integral metrics such as the total distance of exercise, current speed and total ascent of the exercise. The second segment is devoted to the interval information. The three frames in the second segment symbolize past, current and the next intervals: the first frame is a stopwatch, the second frame current HR, while the last frame is an interval stopwatch (that decrements the time). The fourth, or a last segment, is intended for real-time feedback control of the interval. Here, the symbol "-" denotes an athlete performing below the plan, "=" symbolizes that interval right on track, while a "+" athlete performed the interval better than planned.

Athletes interact with the GUI using a touchscreen. Four buttons are intended for full control during the workout. The first button (named "Start tracking") starts the processes of (1) monitoring the sport activity and (2) storing the data in a database, while the second button "Stop tracking" stops both processes. When the athlete is willing to perform the next planned interval, he/she presses the button "START INTERVAL" which starts decrementing the planned time of the interval. The button "STOP INTERVAL" is a reserved one, intended to be pressed only in the case that the athlete can not finish the interval fully. In such case, the AST-Monitor still considers the intermediate results achieved, and stores the part-interval results in the database for post-hoc analysis. An additional (status) button named "POWEROFF" is attached, intended for safe shut down of the AST-Monitor.

The experimental simulations were conducted as follows: A cyclist (man) rode the bicycle equipped with the AST-Monitor on a set of indoor track-rollers. At first, he loaded the training plan proposed by the AST and warmed up. Afterwards, he started to conduct the first interval training session. During the training session, the cyclist monitored his achievement constantly and adapted the effort accordingly. For instance, if the current interval was colored in red, this was a sign that the cyclist needed to increase the effort.

After finishing the intensity phase, the cyclist took a short break in a rest phase, where the decision on the duration of the rest was left with the cyclist. When he felt ready for the next round of the interval session, he pressed the button to start the interval. With such increments, the AST-Monitor guided the cyclist through the entire interval training session, saving him the effort of calculating the deviation from plan manually and not focusing on cycling. Due to the excellent connectivity of the Raspberry Pi, the data monitored and stored were immediately available to the AST after each interval.

In summary, the illustrative example shows that the AST-Monitor supplements the AST successfully in the realization phase of the sport training. On the one hand, it introduces the new automatic measuring methods, while on the other, it is fully dependent on the central AST. Indeed, the AST-Monitor represents an extended arm of the AST capable of retrieving reliable and accurate data in real-time. Surely, some limitations come into play when using this device, like: the size of the device, the size of the display (important for designing the GUI) and the power-bank life.

\section{Conclusion}\label{sec:5}
Although the AST enables automation of all phases of sport training, it is embedded into the central (personal or server) computer, and is therefore not able to monitor the actual performance of an athlete during the training sessions. To overcome this problem, an AST-Monitor is proposed in the paper that represents a wearable Raspberry Pi computer. This device is mounted on the bicycle, and is able to monitor a palette of sensors, measure different training load indicators and communicate with the AST.

The illustrative example using the proposed device during the interval training sessions in the simulated environment showed that this device is capable of measuring a large number of different training indicators. The device is able to track the control phase of the sport training and was thus warmly greeted by the cyclists that tested it. Although only two sport training indicators were used in our study, the emergence of various sensors (like Powermeters) and their simple connectivity to the Raspberry Pi, seems to ensure a very bright future for this device.

There are many directions for the future development, of which (1) connecting the new sensors to the Raspberry Pi, (2) improving the GUI for interacting with the users, (3) automating data transfer to the central (personal) AST computer from the terrain (via a mobile data connection), (4) calling for a social note of the AST-Monitor, either by live tracking or separately, and competing with other cyclists (possible integration with the sport social network Strava), and (5) downsizing the current AST-Monitor.

\end{document}